\documentstyle [12pt] {article}
\begin {document}
\large
\begin {center}
\par
{\bf Possible Development of the Newton Gravitational Theory of
Interactions. An Alternative Approach to the Gravitational Theory}\\
\hspace{0.2cm}

\par
    Kh.M. Beshtoev
\par
Joint Institute for Nuclear Research, Joliot Curie 6, 141980
Dubna, Moscow region, Russia, beshtoev@cv.jinr.dubna.ru
\end {center}

\par
{\bf Abstract} \\

\par
This work is devoted to the discussion of an idea that gravitational
interactions might be residual interactions of strong and electromagnetic
interactions. Then, absence of the carriers of the gravitational interactions
finds a natural explanation in the framework this idea. Besides, since masses
(charges of the gravitational interactions) of particles are generated
in strong, electromagnetic (and possibly, in other) interactions and if
masses of the particles will be not generated in these interactions
(i.e. $m \equiv 0$), then the gravitational interactions do not appear.
That is also indirect confirmation of the considered idea. Connections
between charge of the gravitational and other interactions are considered.\\

\section {Introduction}

The Einstein theory of gravitational interactions [1] supposes that all
particles and bodies having energy, participate in the gravitational
interactions (i.e. gravitational charge is $ m_{eff} = \frac{E}{c^2}, c$
is the light velocity).
The Newton theory of gravity states that only massive particles and bodies
(masses are charges) participate in interactions:
$$
E = - G \frac{m_1 m_2}{r} ,
\eqno(1)
$$
where $E, G$ are energy and constant of gravitational interactions, $m_1,
m_2$ are masses of interacting particles or bodies and $r$ is distance
between bodies or particles. Here, we see that massless bodies and
particles cannot participate in Newton gravitational interactions (photon is
such particle). In the Einstein gravity theory photons participate in the
interactions since they have energy (but $m = 0$) and effective mass is
$m_{eff} = \frac{E}{c^2}$. The Einstein gravity theory is constructed as a
development of the Newton gravity theory and there is supposed that charge of
the gravitational interactions is connected with the forth component of
energy-momenta tensor.  Afterwards, this theory was checked on
practice [2, 3]. But concrete form of interpretation of these experiments is
under critical analysis [4-9], and it is hard to come to conclusion that this
theory is confirmed. In this connection, there arises necessity of searching
(constructing) of the correct theory of gravity, as well as deeper
understanding of the nature of origin of the gravity interactions.

\section {Alternative Approach to the Theory of Gravitational Interactions}

In the work [9] it was shown that in the Newton gravity theory takes place
a single shift of atomic and nuclear levels connected to changing effective
masses of electrons and nucleons in external gravity fields (similar to the
Stark effect [10] in external electric fields or level displacements in external
magnetic fields [11]). And in the Einstein gravity theory [1] since the
photons participate in gravitational interactions there must arise the same
shift at moving photons through the gravitational field, and as a result
there must take place double shift. In the experimental results obtained in
[2, 3], there is only single shift, i.e. the change of photon frequency at
his going through gravitational field does not arise. It means that the
hypothesis that gravitational charge can be connected to the tensor energy-
momentum is not confirmed. Then, the following question arises: Could we
come to this experimental result from the general positions of the theory of
interactions? At present, we know four interaction types: electromagnetic,
strong, weak and gravitational [12]. The theory of electromagnetic, strong
and weak interactions are constructed on the principle of the gauge
invariance. In these theories the charges are scalars and invariant
values (i.e. are not changed depending of coordinate system movements).
In the Newton gravitational theory, the masses are charges, which are
scalars and invariant values, i.e. this theory fulfills the basic demand of the
theory of interactions (remark about gauge invariance of gravity theory, see
below). In the Einstein theory of gravity the demand of invariance
of charge is not fulfilled. Really,  on the experiments on
measurement frequencies shift the Einstein hypothesis is not
confirmed. If the gravitational interactions take place through
forth component of energy-momentum then it is possible to connect
this interaction with  space curvature [1]. Since photons and
another massless particles do not participate in the gravitational
interactions, then necessity in this approach disappears and it is
not necessary to construct the gravity theory on base of the
second rank tensor. Then is stayed open the question about what
types is the gravitational interactions? Obviously, the more attractive
solution in this case is to construct vector theory of gravitation.
\par
An another very interesting approach to the problem of gravity is discussed
in work [13].
\par
What do we know about the gravity? The Newton theory of gravity is well
proved. These interactions are the long distance ones. The sign of gravity
charge $m$ (mass) of particle and antiparticle coincide, i.e. $C$ (charge)
parity is completely violated and the charge of particle and antiparticle
coincide (it means that in these interactions polarization is absent). The
carries of these interactions were not found. Probably it means that there
are no any carriers of these interactions. Besides, masses of particles are
generated mainly by strong and electromagnetic interactions (probably,
there is another source of masses besides the strong and electromagnetic
interactions which generate $W, Z$ boson masses [12]). If the particle
masses are not generated by these interactions, (i.e. $m = 0$) there
cannot be any gravitational interactions. Then, probably, gravitational
interactions are residual interactions of the above interactions (by analogy
with the Van der Vaalse forces [14] in the atomic physics) which generate
masses (charges) of the gravitational interactions.
\par
It is clear that if we want to construct gauge theory of
gravitational interactions it must be $U(1)$ theory since these
interactions are long distance interactions. In $U(1)$ gauge
theory charge must be discrete (quantizated) but in these interactions charge
(masses are not discrete values). And there must be charges with
opposite signs where charges of the same sign are repulsed (and
charges of the opposite sign are attracted). In the gravitational
interactions are only charges (masses) of the same sign and they
are attracted. So we see that it is impossible to construct gauge
theory of the gravitational interactions in a straight manner and probably
these interactions are residual ones with violated $U(1)$ gauge symmetry.
\par
Let us connect the charge of the residual (gravitational) interactions to the
charge of strong interaction. Mass $M_A$ of a nucleus $A$ in rough
approximation is equal to the masses of $Z$ protons and $N$ neutrons
forming this nucleus
$$
M_A \cong Z  m_p + N m_n .
\eqno(2)
$$
Supposing that $m_p \cong  m_n \cong m_N$ ($N$ is nucleon) and the
mass of nucleon is formed  by the strong interactions, one can write the
following expression:
$$
m_N = g_{strong} C  \cong  1 \quad GeV ,
\eqno(3)
$$
where $C$ is constant of the scheme where the mass is computed [12].
\par
\noindent
Then
$$
M_A \cong   A m_N ,
\eqno(4)
$$
and then we can construct gravitational interactions of the nuclei (and
matter composing from nuclei).
Now, we will return to the nucleon. From the Exp.(3) and supposing that
$g_{strong}^2 \cong 0.1$, we have
$$
C^2 = \frac{m_N^2}{g_{strong}^2} \cong 10 \quad GeV^2 .
$$
Now, let's consider energy of gravitational interaction of two nucleons:
$$
E_N = - G \frac{m_N m_N}{r} \cong G \frac{C^2 g_{strong}^2}{r}
= - \bar G \frac{g_{strong}^2}{r}  ,
\eqno(5)
$$
where constant $\bar G$ is the value having no measurement and it is
measure of suppression of the constant of the strong interactions
$$
\bar G C^2 \cong G \cdot 10 \cdot GeV^2 = 1.66 \cdot10^{-4} ,
\eqno(6)
$$
i.e.
$$
E_N \cong - 1.66 \cdot 10^{-4} \frac{g^2_{strong}}{r} = - \bar G
\frac{g^2_{strong}}{r} , \quad g_{grav}^2 = 1.66 \cdot 10^{-4} g_{strong}^2 .
\eqno(7)
$$
We see that in this approach the constant of the gravitational
interactions is a value without measurement and $\bar G$ is measure of suppression
of the strong interactions and residual charge is $g_{grav}$.
\par
Repeating the same consideration taking into account the electric charge
$e$, we get
$$
E_N = - 2.27 \cdot 10^{-3} \frac{e^2}{r} = - \bar G' \frac{e^2}{r} ,\quad
g_{grav}^2 = 2.27 \cdot 10^{-3} e^2 .
\eqno(8)
$$
Probably, the residual charge of strong and electrical interactions must be
equal if the gravitational interactions are universal ones; then the residual
gravitational interactions can appear at a joint of strong and
electromagnetic interactions.
\par
The objective of this work is to attract opinion to the considered possibility,
which explains absence of carriers of the gravitational interactions, and in
this case the couple constant of gravitational interactions is a value without
measurement.

\section {Conclusion}

In this work, we have discussed the idea that gravitational interactions may
be residual interactions of the strong and electromagnetic interactions.
Then, absence of the carriers of the gravitational interactions finds a
natural explanation. Besides, since masses (charges of the gravitational
interactions) of particles are generated in strong, electromagnetic (and
possibly in other) interactions, and if masses of the particles are not
generated in these interactions (i.e. $m \equiv 0$), then the gravitational
interactions do not appear. That is also indirect confirmation of the
considered idea. Connections between the charge of gravitational and other
interactions are considered.
\par
In conclusion we would like to do some remarks about verification of
the gravitational theory:
\par
1. In the experiments with the Sun participant it is necessary to take into
account diffraction, refraction and the strict content of the Sun atmosphere.
\par
2. In the experiments with the light and the radio wave in the Sun
system it is necessary to take into consideration strict compositions
and distributions of gasses in the Sun system.
\par
3. In the experiments on linsing it is also necessary to take into
account strict composition and distribution of gasses in inter
star and inter galactic space.\\

\par
{\bf References}\\

\par
\noindent
1. A. Einstein, Ann. Phys. (Leipzig) {\bf 49}, 769, (1916);
\par
 L.D. Landau, E.M. Lifshits, Field Theory, M., Nauka,
\par
1988, p.324.
\par
\noindent
2.  R.V. Pound,  G.A. Rebka, Phys. Rev. Let. {\bf 4}, 337, (1960);
\par
 R.V. Pound, J.L. Snider, Phys. Rev. {\bf 140}, 788, (1965).
\par
\noindent
3.  J.L. Snider, Phys. Rev. Let. {\bf 28}, 853, (1972).
\par
\noindent
4. P. Marmet, Einstein's Theory of Relativity versus Classic
\par
Mechanics, Newton Physics Books, Canada, 1997;
\par
\noindent
5. P.  Marmet and C. Couture, Physics Essays, 1999, v.12, p.162.
\par
\noindent
6. V.N. Strel'tsov, JINR Communic. P2-96-435, Dubna, 1996; JINR
\par
Communic. P2-98-300, Dubna, 1998; Apeiron, {\bf 6}, 55,(1999).
\par
\noindent
7. V.V. Okorokov, ITEP preprint N 27, Moscow, 1998.
\par
\noindent
8. L.B. Okun, K.G. Selivanov and V.L. Telegdi, UFN (Russian
\par
Journ.)  169, 1140, (1999).
\par
\noindent
9.  Kh. M. Beshtoev, JINR Commun. P4-2000-45, Dubna, 2000;
\par
Physics Essays 2000, v.13, p.593; Procced. of the 27-th ICRC,
\par
Hamburg, Germany, v.3, p.1183.
\par
\noindent
10. G. Bethe, E. Solpiter, Quantum Mechanics of Atoms
\par
with One and Two Electrons, Moscow, 1960.
\par
\noindent 11. W. Gerlach, O. Stern, Der Experimentelle Nachwies
\par
der Richtungsquantelung in Magnetfield, Z. Phys.,
\par
1922, Bd.9, s.349.
\par
\noindent
12. The Europian Phys. Journ. C, Review of Particle Phys. 2000,
\par
v.15, N1-4.
\par
\noindent
13. A. A. Tayapkin, Proceedings of the IV-th Scientific
\par
Conference Young Scientists and Specialists of JINR, Dubna,
\par
2000, p.49.
\par
\noindent
14. A. L. Buchachenko, Chimical Polarizations of Electrons
\par
and Niclei, Moscow, 1974.

\end{document}